\documentclass[11pt,a4paper]{article}
\begin{document}

\title{ The tortoise coordinates and the cauchy problem in the stable study of
 the Schwarzschild black hole  \footnote{ E-mail of Tian:
 hua2007@126.com, tgh-2000@263.net}}
\author{Gui-hua Tian,\ \ Shi-kun Wang,  \ \ Shuquan Zhong\\
School of Science, Beijing University \\
of Posts And Telecommunications, Beijing 100876, P.R. China\\
Academy of Mathematics and Systems Science,\\ Chinese Academy of
Sciences (CAS), Beijing 100080, P.R. China }
\date{}
\maketitle

\begin{abstract}
Generally, the Schwarzschild black hole was proved stable through
two different methods: the mode-decomposition method and the
integral method. In the paper, we show the integral method can
only apply to the initial data vanishing at both the horizon and
the spatial infinity. It can not treat the initial data only
vanishing at the spatial infinity. We give an example to show the
misleading information caused by the use of the tortoise
coordinates in the perturbation equations. Subsequently, the
perturbation equation in the Schwarzschild coordinates is shown
not sufficient for the stable study.

\end{abstract}
\section*{1. Introduction}
In general relativity, there are two famous stationary black
holes: the Schwarzschild black hole and the Kerr black hole. The
Schwarzschild black hole comes from complete gravitational
collapse of the massive spherically symmetric star, while the Kerr
black hole results from the complete gravitational collapse of the
massive spinning star. Now, the stable properties of these two
black holes are crucial theoretically and astronomically in
general relativity . When the black hole is proved stable, it
could be candidate for the gravitational wave source. The
Earth-based interferometers include LIGO, Virgo, GEO600 and TAMA,
and the space-based interferometer is LISA. These interferometers
all aim at the detection of the gravitational wave. The promising
sources of the gravitational wave are the inspiring binaries black
holes. Among these sources, the extreme-mass-ratio binaries are
primary ones. The extreme-mass-ratio binary could be modelled by
perturbation of a massive black hole with mass $M$ by a small body
with mass $m$. $\frac m M$ ranges from $10^{-1}$ to $10^{-9}$,
though $m$ could be as large as $100M_{\odot}$. What kinds of
gravitational fields could be candidate for the massive black
hole? Only those whose gravitational fields are stable be
qualified. Therefore, the stable study of the black holes becomes
urgent now.

It is generally believed that the stable problem of the
Schwarzschild black hole has been settled, while the stable
properties of the Kerr black hole remain controversial and
unsolved.

We have restudied the stable problem of the Schwarzschild black
hole and found that its stable properties are still unsolved :
they really depend on the time coordinate and where the initial
time slice intersects the horizon\cite{tian1}-\cite{tian4}.

Originally, it is in the Schwarzschild coordinate system
\begin{equation}
ds^{2}=-(1-\frac{2m}{r})dt^{2}+(1-\frac{2m}{r})^{-1}dr^{2}+r^{2} d
\Omega ^{2},\label{orimetric}
\end{equation}
where the stable problem is studied. The well-known perturbation
equations are
\begin{equation}
\frac{\partial^{2}Q}{\partial t^2}-\frac{\partial^{2}Q}{\partial
r^{*2}}+VQ=0,\label{rw}
\end{equation}
is obtained in the very coordinates. In the perturbation equations
(\ref{rw}), the tortoise coordinate $r^*$ is defined by
\begin{equation}
r^{*}=r+2m\ln \left(\frac{r}{2m}-1\right)\label{tortoise}
\end{equation}
which approaches $-\infty$ at the horizon $r=2m$ and $\infty$ at
spatial infinity, and the effective potential $V$
\begin{equation}
V=\left(1-\frac{2m}{r}\right)\left[\frac{l\left(l+1\right)}{r^{2}}+\frac{2m(1-s^2)}{r^{3}}\right]
\label{potential}
\end{equation}
is positive over $r^*$ from $-\infty$ to $\infty$. $s$ could take
the value $0$, $1$, $2$ with the Eq.(\ref{rw}) corresponding to
the scalar, electromagnetic and gravitational perturbation
equations respectively. Actually, the Eq.(\ref{rw}) becomes the
well-known Regge-Wheeler equation when $s=2$.

Generally, researchers treated the stable problem of Eq.(\ref{rw})
by mode-decomposition method. Because the Eq.(\ref{rw}) are
written by the the  Schwarzschild time coordinate $t$ and $r$ or
$r_*$, the time coordinate $t$ is taken for granted to define the
initial time. Then the  Schwarzschild black hole was proven
stable\cite{vish}-\cite{wald1}.

Later, the integral method appeared. Researchers also investigated
the stable problem by the method. In this way, they proved that
any initially compact-supported perturbation data (relative to the
tortoise coordinate $r_*$) can not blow up in the later
time\cite{chan},\cite{wald}. This method confirmedly put an end to
any doubts about the stable property of the Schwarzschild black
hole .

We first noticed that there exists some flaws in the
mode-decomposition method. Because the metric component $g_{00}$
is zero at the horizon $r=2m$ in the  Schwarzschild coordinates,
the Schwarzschild time coordinate $t$ loses its meaning at the
horizon. Actually, the future horizon corresponds to $t\rightarrow
\infty $, and the past horizon corresponds to $t\rightarrow
-\infty $. This shows that the coordinates $(t,\ r)$ do not cover
the horizons. So, $t$ is not suitable to define the initial time
containing the horizons.

The Kruskal coordinates cover the whole horizons of the
Schwarzschild black hole. The two Painlev\'{e} coordinates cover
the future and past horizon respectively. Correspondingly, these
time coordinates are good for the investigation of the stable
problem. If we use these "good" time coordinates to define the
initial time, the stable properties depend complicatedly on where
the initial time slice intersects with the black hole: the
Schwarzschild black hole is stable when the initial time slice
intersects the future horizon of the black hole, whereas the
Schwarzschild black hole is unstable when the initial time slice
intersects the past horizon of the black
hole\cite{tian1}-\cite{tian4}. These extraordinary results are in
striking contrast with the conclusion taken for granted that the
Schwarzschild black hole is stable. They make people puzzled on
the problem.

Afterward, we study the stable problem of the Rindler space time
for finding the real answer to the Schwarzschild black hole's
stable problem\cite{tian6R}. The Rindler space time has the much
similar or almost the same geometrical structure with the
Schwarzschild black-hole, and is mathematically simple for study
of the stable problem. It is found that the same situation exists
for the Rindler space time if we study the scalar field equation
completely in the Rindler space time: its stable properties also
depends complicatedly on where the initial time slice intersects
with the Rindler space time. The Rindler space time is stable when
the initial time slice intersects its future horizon, whereas it
is unstable when the initial time slice intersects its the past
horizon \cite{tian6R}.

Unlike the Schwarzschild black hole, the Rindler space time is one
part of the Minkowski space time, which is the most suitable and
simplest space time to study. We could study the scalar field
equation completely in the Minkowski space time coordinate system
and obtain definite answer to the stable problem of the Rindler
space time: in this very way, it has been found that the Rindler
space time is unstable\cite{tian6R}. This answer clearly shows
which is correct among the controversial answers to the Rindler
space time's stable problem, that is, the answer whose initial
time slice intersects the past horizon is correct
answer\cite{tian6R}. On the other hand, the Minkowski space time
is stable, so the Rindler space time could exist as one part of
the Minkowski space time.

Though we have not studied the perturbation equation completely in
Kruskal coordinate system, which is  too difficult to study, we
still could partly infer and partly guess  that the Schwarzschild
black hole might be unstable by comparison with the case of the
Rindler space time. We also infer that the Kruskal space time
might have real entity and the Schwarzschild black hole could
exist stably only as one part of the Kruskal space time.

If we insist that the stable property of the Schwarzschild black
hole is not settled, we must find out why the integral method is
wrong. This is main purpose of the paper.

In the paper, we show the integral method is wrong for the initial
data not vanishing at the horizon in section 2. We give an example
in section 3 to show the effect of the tortoise coordinates on
stable study. In section 4, we also show that the perturbation
equations in the Schwarzschild coordinates (tortoise coordinates )
might lose some information. They wrongly lead one to believe that
the wave function inside the black hole ($r<2m$) could not
influence the region outside of the black hole ($r>2m$). So, the
perturbation equations in the Schwarzschild coordinates is not
sufficient for the stable study.

\section*{2. Misleading concept of compact property by the tortoise coordinates}

Why researchers obtained that the Schwarzschild black hole is
stable by the integral method? In the paper, we show that the
tortoise properties of the Schwarzschild coordinates $t$ and $r_*$
are responsible for that.

Generally, the stable problem is believed to be settled by much
more mathematically rigorous proof  based on the  integral method
\cite{vish}-\cite{wald1}.

The perturbation equations (\ref{rw}) could be written as
\begin{equation}
\frac{\partial^{2}Q}{\partial t^2}=\frac{\partial^{2}Q}{\partial
r_*^{2}}-VQ=0,\label{rw2}
\end{equation}
then the operator
\begin{equation}
A=-\frac{d^{2}}{dr_*^{2}}+V,\label{operator}
\end{equation}
is positive and self-conjugate on the Hilbert space $L^2(r_*)$ of
square integral functions of $r_*$ . It is mathematically by these
very properties of the operator $A$ and the compact-supported-ness
of the quantity $Q$ that the Schwarzschild black hole is proven
stable\cite{chan}-\cite{wald}. The mathematical method is
generally believed correct.

The main problem in references lies on the use of the misleading
concept "compact-supported-ness of the quantity $Q$": this concept
is used in the tortoise coordinates $t$ and $r_*$. Actually, $r_*$
is the spatial tortoise coordinate, which changes the horizon of
the Schwarzschild black hole $r=2m $ finite into negative infinity
by $r_*$. It is well-known that the horizon really lies in finite
spatial proper distance from some point $r=r_0>2m$. The horizon
corresponding to $r_*\rightarrow -\infty$ is only the effect of
the tortoise coordinate $r_*$.

Really, the integral proof sounds reasonable, but it has the flaw
in using the compact-supported property relative to the tortoise
coordinate $r_*$. In fact, there are two kind of initial data. The
first one vanishes both at the spatial infinity and at the
horizon, which usually is denoted by the compact-supported initial
data and used in the literatures. This data is really
compact-supported relative to both the coordinate $r$ and the
tortoise coordinate $r_*$. Because the horizon truly locates at
finite position by the coordinate $r$, the initial data need not
be zero at the horizon. There exists the second kind of the
initial data, which approaches zero only at the spatial infinity.
The second initial data is compact-supported only relative to the
coordinate $r$. Therefore, only confined to the first kind of
initial data, the integral method is right.

Furthermore, the Schwarzschild time coordinate $t$ is really also
a tortoise coordinate. It is also well-known that a traveller only
needs finite proper time to arrive at the horizon of the
Schwarzschild black hole and fall into it. In contrast to the
proper time, it takes the tortoise coordinate $t$ to infinity for
the traveller to arrive at the horizon of the Schwarzschild black
hole. This really means that any finite proper time process
containing the horizon of the Schwarzschild black hole corresponds
to $t\rightarrow \infty$ of the Schwarzschild time coordinate.

The tortoise property of the Schwarzschild time coordinate $t$
makes it not suitable for the stable study at the horizon: for
example, it may only take short proper time for some physical
fields to arrive at the horizon, while it corresponds to the
Schwarzschild time coordinate $t\rightarrow \infty$. In this case,
the compact-supported-ness of the physical quantity $Q$ with
respect to the tortoise coordinates $t$ and $r_*$ are really
misleading.

For example, the positive property of the operator $A$ of
Eq.(\ref{operator}) relies on the compact-supported-ness of the
quantity $Q$. From Eq.(\ref{rw2}), one obtains the following
equation
\begin{equation}
\frac{\partial}{\partial t}\left[\int_{-\infty}^{+\infty}\mid
\dot{Q}\mid
^{2}dr_*+\int_{-\infty}^{+\infty}Q^*AQdr_*\right]=g(t)\label{g(t)}
\end{equation}
where $g(t)$
\begin{equation}
g(t)=-\lim_{R\rightarrow \infty}\left[Q^*\frac{\partial
\dot{Q}}{\partial r_*} -\dot{Q}\frac{\partial Q^*}{\partial r_*}
\right]_{-R}^{+R},\ R\rightarrow \infty.\label{g(t)2}
\end{equation}

When $g(t)=0$ in Eq.(\ref{g(t)}), the integral method proves that
the quantity
\begin{equation}
\left[\int_{-\infty}^{+\infty}\mid \dot{Q}\mid
^{2}dr_*+\int_{-\infty}^{+\infty}Q^*AQdr_*\right] \label{integral
const}
\end{equation}
is constant, this in turn excludes the blowing up of the
perturbation fields $Q$. Then, the black hole is proved stable
\cite{chan}-\cite{wald}. But $g(t)$ can be non-zero. As it is just
stated, the real physical process arrives at the horizon when
$t\rightarrow \infty $, though it takes only finite proper time
actually. So, we must study the process at the tortoise time
$t=\infty +\bar{t}$ where $\bar{t}$ is finite. Similarly, $t$ in
the Eq.(\ref{g(t)}) is also replaced by $\infty +\bar{t}$.
Furthermore, the quantity $Q$ could be non-zero at the horizon. Of
course, it vanishes initially at the spatial infinity. That is
$Q=0$ at $r\rightarrow \infty$ or $r_*\rightarrow \infty$, and
$Q\ne 0$ at $r=2m$ or $r_*\rightarrow -\infty$. In this case,
$g(t)$ is generally no longer zero. The mode-decomposition method
is not suitable in the Eq.(\ref{g(t)}), nevertheless, we use the
mode-decomposable solution $Q_{k}(r_{*})e^{-ikt}$ as an example
with $Q_{k}(r_*)=B_k e^{ ikr_{*}}$ when $r_{*}\rightarrow -
\infty$. Actually by the Eq.(\ref{g(t)2}),
$g(t)=2|B_{k}|^{2}|k|^{2}> 0$ is obtained for the solution. Hence
the mathematical proof in references fails.

Therefore, the Reege-Wheeler equation is not suitable for
mathematical proof involving the concept of the
compact-supported-ness of the physical quantities. The stable
proof based on this very concept is actually false, no matter how
it seems mathematically rigorous.

\section*{3. The Rindler case as an example}

In the following, we use the Rindler space time as an example to
show why the compact-supported-ness is not suitable in the
tortoise coordinates. Suppose we solve the scalar field for the
Rindler space time. Because the Rindler space time is one part of
the Minkowski space time, we could write the equation in the
Minkowski coordinates. The Minkowski space time's metric is
\begin{equation}
ds^{2}=-dT^{2}+dZ^{2}+dx^{2}+ dy^{2},\label{orimetric-m}
\end{equation}
and the scalar field equation in the Minkowski space time is
\begin{equation}
-\frac{\partial^{2}\Psi}{\partial T^{2}}+ \frac {\partial ^2
\Psi}{\partial Z^2}+ \frac{\partial^{2}\Psi}{\partial
x^{2}}+\frac{\partial^{2}\Psi}{\partial
y^{2}}-\mu^{2}\Psi=0.\label{kg in m}
\end{equation}
The Rindler space time corresponds to the part $Z>0,\ Z^2-T^2>0$
of the Minkowski space time. $Z^2-T^2=0$ corresponds to the
horizon of the Rindler space time. The boundaries for the
Eq.(\ref{kg in m}) consist of the horizon $Z^2-T^2=0$ and the
spatial infinity $Z\rightarrow \infty$.

The dependence of $\Psi$ on the variable $x,\ y$ could be
\begin{equation}
\Psi =\psi(Z,T)e^{ik_1x+ik_2y}
\end{equation}
where $\psi$ satisfies
\begin{equation}
\frac{\partial^2\psi}{\partial T^2}= \frac {\partial ^2
\psi}{\partial Z^2}- (k_1^2+k_2^2+\mu^2)\psi.\label{kg in m2}
\end{equation}
The operator $A_1$ is defined as
\begin{equation}
A_1= -\frac {d^2 }{dZ^2}+ (k_1^2+k_2^2+\mu^2).\label{operator a1}
\end{equation}
Corresponding to the Minkowski coordinates $T$ and $Z$, we could
use the concept of the compact-supported-ness of the $\psi$, that
is, the initial compact data of $\Psi$ will be compact at finite
time $T$. So we could obtain $\psi (T,Z)=0$ as $Z\rightarrow
\infty$. Because the horizon is $Z^2-T^2=0$, generally, $\psi
(T,Z)\ne 0$ at $T$. For the operator $A_1$, we see that
\begin{eqnarray}
I(\psi )&=& \int_{horizon}^{+\infty}\psi^*A_1\psi
dZ=\int_{horizon}^{+\infty} \left[-\psi^*\frac {\partial^2\psi
}{\partial Z^2}+ (k_1^2+k_2^2+\mu^2)\psi
\psi^*\right]dZ\nonumber \\
&=& \left[-\psi^*\frac {\partial \psi }{\partial
Z}\right]_{horizon}^{\infty}+ \int_{horizon}^{+\infty} \left[\frac
{\partial \psi^* }{dZ}\frac {\partial \psi }{dZ}+
(k_1^2+k_2^2+\mu^2)\psi \psi^*\right]dZ. \label{a3}
\end{eqnarray}
Because the first term in Eq.(\ref{a3}) could be negative for some
$\psi$, the operator $A_1$ is no longer a positive operator.

For example, $I(B_0(T)e^{-kZ})=\frac
{|B_0(T)|^2}{2k}\left[(k_1^2+k_2^2+\mu^2)-k^2\right]e^{-2kZ_0}$
could be negative for some $k>\sqrt{k_1^2+k_2^2+\mu^2}$ with
$Z_0=\pm T$. This again shows that the Rindler space time is not
stable.

On the other hand, we could write the scalar equation in the
Rindler coordinates. The Rindler metric is
\begin{equation}
ds^{2}=-(1+az)dt^{2}+(1+az)^{-1}dz^{2}+dx^{2}+
dy^{2},\label{orimetric r}
\end{equation}
and the horizon is at $z=-\frac 1a$. The coordinates  $t,\ z$ are
related with $T,\ Z$ by
\begin{equation}
T=\frac2a\sqrt{1+az}\sinh{\frac{at}2},\  \
Z=\frac2a\sqrt{1+az}\cosh{\frac{at}2}.
\end{equation}
The scalar equation in the Rindler coordinates is
\begin{equation}
-\frac1{1+az}\frac{\partial^2\psi}{\partial t^2}+ \frac {\partial
}{\partial z}\left[(1+az)\frac {\partial \psi }{\partial
z}\right]- (k_1^2+k_2^2+\mu^2)\psi=0.\label{kg in r2}
\end{equation}
Define the spatial tortoise coordinate $z_*$ as
\begin{equation}
z_*=\frac 1a\ln(1+az)
\end{equation}
which makes the horizon $1+az=0$  and the spatial infinity into
$z_*\rightarrow -\infty$ and $z_*\rightarrow +\infty$
respectively. The Eq.(\ref{kg in r2}) then becomes
\begin{equation}
\frac{\partial^2\psi}{\partial t^2}- \frac {\partial ^2
\psi}{\partial z_*^2}+(1+az) (k_1^2+k_2^2+\mu^2)\psi =0.\label{kg
in r3}
\end{equation}
Then the operator $A_2$ is
\begin{equation}
A_2= -\frac {d^2 }{dz_*^2}+
(1+az)(k_1^2+k_2^2+\mu^2).\label{operator a2}
\end{equation}
The Eqs.(\ref{kg in r3}) , (\ref{operator a2}) for the scalar
field in the Rindler space time are almost the same as that in the
Schwarzschild black hole (The only difference lies in the spatial
infinity where the effective potential
$V=(1+az)(k_1^2+k_2^2+\mu^2)$ does not fall off to zero in the
Rindler space time).

From above discussion, the scalar field is not compact in the
tortoise coordinates $t,\ z_*$ because $\psi$ could be non-zero at
the "good" finite time $T$ on the horizons $Z^2-T^2=0$. But the
Eqs.(\ref{kg in r3}) , (\ref{operator a2}) and the Rindler metric
(\ref{orimetric r}) can easily mislead one the wrong concept that
$\psi$ should be compact. Subsequently, one might wrongly prove
that the Rindler space time is stable with respect to the scalar
perturbation.

In the above example, we see that it is the tortoise coordinates
that make us more liable to error and this error may be much more
elusive  to find. This tortoise property of the Schwarzschild time
coordinate $t$ has been not noticed by researchers in stable
study, though it has been known in other aspects study. Actually,
the influence of the tortoise property of $t$ is much more obscure
and less noticed than that of $r_*$.

\section*{4. Phase velocity and the cauchy problem}

The Eq.(\ref{kg in r3}) has another  blemish that is illusive in
the stable investigation of the Rindler space time. From it, we
see even that the phase velocity in the direction of $z$ axis of
the scalar field can not be faster than that of the light at the
future horizon. Whether or not its mass $\mu $ and wave length in
other direction $k_1,\ k_2$ are zero, the the phase velocity in
the direction of $z$ axis is $1$ at the horizons. The fact  really
makes people regard the Eq.(\ref{kg in r3}) and the initial data
on the hypersurface $t=const$ are a well-posed Cauchy problem in
classical relativity. This very fact comes from the use of the
tortoise coordinates, and is illusive. Actually, it means the
scalar field in the region $Z^2-T^2<0, \ T>0$ can not have any
influence on the Rindler space time. This is also wrong.

Though the velocity of any classical particle can not be greater
than that of the light, the wave phase velocity could exceed speed
limit. From the Eqs.(\ref{kg in m}), (\ref{kg in m2}), we easily
see that the phase velocity in the direction of $Z$ axis could be
greater than that the velocity of the light at the future horizon.
So, the scalar field in the region $Z^2-T^2<0, \ T>0$ definitely
has influence on the Rindler space time.

Similarly, we give the massive scalar perturbation equation as an
example. The scalar perturbation equation in the Schwarzschild
coordinates is
\begin{equation}
\frac{\partial^{2}Q}{\partial t^2}-\frac{\partial^{2}Q}{\partial
r^{*2}}+\left(1-\frac{2m}{r}\right)\left[\frac{l\left(l+1\right)}{r^{2}}
+\frac{2m}{r^{3}}+\mu^2\right]Q=0\label{rwscalar}
\end{equation}
where $\mu$ is its mass. By the geometric approximation, the
scalar wave $Q$ has the asymptotic forms
\begin{equation}
Q=Ae^{i\omega t\pm ikr^{*}}\label{q-2m}
\end{equation}
with $\omega=k$ at the horizons. This really means that the phase
velocity
\begin{equation}
v_p=\pm \frac{\omega }{k}=\pm 1\label{v-2m}
\end{equation}
is the same as that of the light at the horizon. Therefore, one
gets the scalar field inside the black hole ($r<2m$) could not
influence the counterpart in the Schwarzschild black hole
($r>2m$). This fact is truly wrong, and caused by the tortoise
coordinates $t,\ r_*$. Rewriting the scalar perturbation equation
in the Kruskal coordinates
\begin{equation}
ds^{2}=\frac{32m^3}{r}e^{-\frac
r{2m}}\left[-dT^{2}+dX^{2}\right]+r^{2} d \Omega
^{2},\label{kruskalmetric}
\end{equation}
and by the decomposition of the variables $Q=\psi (T,X)Y_{lm'}$ ,
where $Y_{lm'}$ are the spherical harmonic functions, we obtain
\begin{equation}
\frac{\partial^{2}\psi}{\partial
T^2}-\frac{\partial^{2}\psi}{\partial
X^{2}}+\left(\frac{32m^3}{r}e^{-\frac
r{2m}}\right)\left[-\frac{T}{2mr}\frac{\partial \psi }{\partial T
}-\frac{X}{2mr}\frac{\partial \psi }{\partial X
}+\mu^2+\frac{l\left(l+1\right)}{r^{2}}\right]\psi=0.\label{scalar
in kruskal}
\end{equation}
By the geometric approximation at the horizon, we obtain the
asymptotic form for $\psi$
\begin{equation}
\psi =A'e^{i\omega T\pm ikX}\label{psi-2m}
\end{equation}
with the relation
\begin{equation}
\omega
^2=k^2+16m^2e^{-1}\left[\mu^2+\frac{l\left(l+1\right)}{4m^2}\right].\label{disper
relation }
\end{equation}
Therefore, the phase velocity at the horizon
\begin{equation}
V_p=\pm \frac{\omega }{k}\label{V-2m}
\end{equation}
is greater than $c=1$ of the light whenever the mass $\mu$ or $l$
is not equal to zero.

Furthermore, the second kind of data was investigated in
Ref.\cite{wald1}, and was used to prove that the Schwarzschild
black hole is stable. In the usual Penrose diagram, part $I$
corresponds to our region($ r>2m $), parts $II$ and $II'$ are the
black-hole($ r<2m $) and white-hole($ r<2m $) regions
respectively, and part $I'$ is another region($ r>2m $) not
communicating with our region. In the proof, it was assumed that
the scalar field in the regions $I'$ and $II$ ($r>2m$) had no
influence on the scalar field studied in the region $I$. From
above discussion, the phase velocity of the Klein-Gordon equation
in the Kruskal coordinates is greater than that of the light even
at the horizon. This in turn influences the propagation of the
scalar fields. So the proof of the theorems in Ref.\cite{wald1} is
not guaranteed (The causal property can not guarantee that
"$\varphi'$" coincides with $\varphi $ in the intersection of the
region I with the future of $C'$ \footnote{see Ref.\cite{wald1}
page 895} ).

The method in references really applies only to perturbations of
the Schwarzschild black hole that vanish at the bifurcation
surface. It can not treat the perturbations with initial non-zero
data at the horizon. In addition, the perturbation equations in
the Schwarzschild coordinates mislead one and are not sufficient
for the stable study.

Therefore the stable properties of the Schwarzschild black-hole
still remain unsettled now, and we only partly infer that it is
unstable by comparison with that in the Rindler space time. On the
other hand, the Kruskal space time might be stable and have
physical entity in comparison with that in the Rindler space time.
So the Schwarzschild black-hole could be stable only as one part
of the Kruskal space time\cite{tian6R}.

\section*{Acknowledgments}

We are supported in part by the National Natural Science
Foundation of China under Grant  Nos.10475013, 10375087, 10373003;
the National Key Basic Research and Development Program of China
under Grant No. 2004CB318000 and the Post-Doctor Foundation of
China.

\end{document}